\documentclass[12pt,preprint]{aastex}
\begin{document}

\title{An International Ultraviolet Explorer Archival Study of Dwarf Novae in Outburst}
\author{Ryan T.Hamilton, Joel A. Urban, Edward M. Sion, Adric R. Riedel, 
        Elysse N. Voyer, John T. Marcy, Sarah L. Lakatos}

\affil{Dept of Astronomy \& Astrophysics,
       Villanova University,
       Villanova, PA 19085,
       e-mail: ryan.hamilton@villanova.edu, jurban@ast.vill.edu, emsion@ast.vill.edu, adric.riedel@villanova.edu, 
               elysse.voyer@villanova.edu, john.marcy@villanova.edu sarah.lakatos@villanova.edu}

\begin{abstract}

We present a synthetic spectral analysis of nearly the entire far 
ultraviolet International Ultraviolet Explorer (IUE) archive of spectra of 
dwarf novae in or near outburst.  The study includes 46 systems of all dwarf nova 
subtypes both above and below the period gap. The spectra were uniformly 
analyzed using synthetic spectral codes for optically thick accretion 
disks and stellar photospheres along with the best-available distance 
measurements or estimates.  We present newly estimated accretion rates and discuss 
the implications of our study for disk accretion physics and CV evolution. 

\end{abstract}
\keywords{Stars: white dwarfs, stars: dwarf novae, accretion disks}

\section{Introduction}

Dwarf novae (DNe) are a subclass of cataclysmic variables (CVs), comprised
of a low-mass, main sequence secondary star and a white dwarf (WD)  
primary. They are
characterized by their quasi-periodic outburst episodes that are typically
2-6 mag in amplitude.   
The Roche lobe-filling secondary loses gas through the inner langrangian point carrying substantial angular momentum that leads to the formation of a disk
around the accreting white dwarf. 
A thermal-viscous instability known as the disk instability model (Osaki 2005 and references therein) causes the accretion disk to transition from a cool, quiescent, optically thin state to a stable state where the disk is much hotter, more luminous, optically thick and approaching a steady state. The disk is heated due
to the release of gravitational potential energy as the accreted material
spirals through the optically thick disk toward the WD surface.  The
energy released can heat the disk to temperatures on the order of 100,000
K.  This heating of the disk accounts for the increase in luminosity
during a dwarf nova (DN) outburst, and should dominate the system's
luminosity in outburst. When the outburst concludes, the disk should,
theoretically, be relatively empty of material in its hotter inner regions
and optically thin.  At this time, the disk luminosity
should be a negligible contributor to the system's luminosity.  This
period of low mass-transfer is known as quiescence.

In dwarf novae, it has long been noted that the systems generally undergo a 
transition from emission line-dominated spectra in the optical and UV (in 
quiescence) to absorption line spectra as the continuum gradually increases
in strength up to outburst maximum with the development of broad absorption
troughs and in the UV, P Cygni profile structure  in the resonance 
lines. In the optical, emission line cores remain but have narrower widths 
than their counterparts in quiescence (Warner 1995 and refererences therein). 
The spectra in outburst and quiescence exhibit a mix of H and He lines as 
well as metals, unlike the dichotomy between the DA (H-rich) and DB (He-rich)
composition sequences of the isolated white dwarfs. 

Two types of dwarf nova outbursts have been recognized, Type A, which have a fast optical rise, delay progessively at shorter wavelengths and describe a wide loop in the two color diagram as the system progresses from quiescence to outburst, then back to quiescence, and Type B, in which the optical rise is slower and  the cycle describes a
narrow loop in the two color diagram with only a small or zero delay between
the UV and optical (Smak 1984a,b) On the decline from outburst, systems of both types show fluxes that fall simultaneously at all wavelengths but with the UV declining fastest near the end of the decline. Pringle \& Verbunt (1984) and Verbunt (1987) 
found that the UV flux distributions evolve similarly for a 
given kind of outburst.

Early attempts to theoretically interpret the spectral energy 
distributions of dwarf novae in outburst relied on comparisons of the the 
ratios of continuum fluxes in emission-line free regions with model 
predictions. The UV flux distributions of DNe in outburst have been mostly 
fitted with $F_{\lambda}\simeq\lambda^{\alpha}$ flux distributions where $\alpha \sim -2$ 
(Verbunt 1987). Moderately good agreement was achieved for systems 
expected to approximate steady state disks. The largest uncertainty in the 
disk model fitting is the sensitivity to the WD mass (Verbunt 1987). The 
outburst spectra have overall flux distributions that resemble in both 
the optical and UV, the flux distribution of a B2-3 V-III star (T$_{eff}\sim20,000$K).

la Dous (1991) noted in her archival study of 32 dwarf novae in outburst 
that the UV continuum flux distributions were virtually identical 
regardless of inclination angle, except for the systems OY Car, Z Cha, AB 
Dra, WW Cet, CM Del, RX And, and VW Vul (with increasingly bluer 
distributions but all redder than the bulk of objects) and UZ Ser that is 
much bluer than the other objects. la Dous (1991) used the published flux 
plots from an earlier generation of accretion disk models to conclude that 
the inner disk radius, inclination angle and mass transfer rate at a fixed WD 
mass of 1 $\rm{M}_{\sun}$ are the most critical factors in modeling the 
observed energy distribution. In particular, the appearence of the UV 
spectra at outburst is strongly correlated with the orbital inclination. 
In lower inclination systems the outburst spectra appear in pure 
absorption while for higher inclinations ($70 < i < 80$ degrees) the spectra 
are continuous and finally for $i > 80$ degrees, strong emission lines are 
present (la Dous 1991).

Accretion rates have been obtained from absolute magnitudes $\rm{M}_V$ 
(requiring distances) for various subgroups of CVs with due account for the 
sensitivity to system parameters such as the inclination angle (Smak, 1989, 1994). 
Generally, during dwarf nova maximum, outburst accretion rates have been estimated 
in the range $\dot{\rm{M}}=3\times10^{-9}$ to $1\times10^{-8} \rm{M}_{\sun}yr^{-1}$ 
while during the rise to outburst among dwarf novae, accretion rates in the range 
$\dot{\rm{M}}=5\times10^{-10}$ to $3\times10^{-9} \rm{M}_{\sun}yr^{-1}$ have been 
derived (Warner 1995). However, relatively few individual accretion rates 
have been published for dwarf novae in outburst based upon detailed 
accretion disk model fitting. There is a comprehensive atlas of optical spectra of 
dwarf novae in outburst (Morales-Rueda and Marsh 2002) but none for UV spectra in which the data has been modeled and parameters derived.  Some examples of early derivations of 
accretion rates in outburst from disk model fitting are for VW Hyi, Z Cha, 
OY Car and SS Cygni. For VW Hyi, an accretion rate 
$\dot{\rm{M}}=5\times10^{-9} \rm{M}_{\sun}yr^{-1}$ was obtained by Polidan, Mauche and 
Wade (1990), and Polidan and Holberg(1984). For Z Cha, Horne and Cook (1995) 
obtained $\dot{\rm{M}}=2.5\times10^{-9} \rm{M}_{\sun}yr^{-1}$ and for OY Car, Rutten 
et al. (1992) obtained $\dot{\rm{M}}=1.6\times10^{-9} \rm{M}_{\sun}yr^{-1}$. For SS Cygni 
at maximum, Kiplinger (1979) found excellent agreement between a model accretion disk 
with $\dot{\rm{M}}=8.5\times10^{-8} \rm{M}_{\sun}yr^{-1}$ and the observed flux 
distribution from the IR through the FUV regions. If $\dot{\rm{M}}$ is not constant 
throughout the disk, then steady state disk model fitting in outburst will yield 
discrepancies due to a non-steady state radial temperature distribution, T(r), in the disk.

Of course, there is a strong sensitivity to the white dwarf mass which is 
poorly known for the majority of systems. Orbital inclinations are 
likewise uncertain and distances remain for the most part rough estimates 
except for less than two dozen systems that have reliable trigonometric 
parallaxes. Thus, even for a given inclination angle and distance, the accretion 
disk fits to be described in section 3 yield inferred white dwarf masses uncertain 
by at least three-tenths of a solar mass. While our spectral coverage is restricted 
and excludes the regions where the boundary layer and the white dwarf probably 
contribute to the FUV flux, the International Ultraviolet Explorer (IUE) spectra do 
cover the range of wavelengths where accretion disks during outburst are emitting a large 
fraction of their light. In view of all these caveats, our disk model fits 
should be regarded only as first approximations to the real accretion 
rates during outburst. Moreover, these spectra nicely complement the 
compilation of optical spectra of dwarf novae in outburst described by 
Morales-Rueda and Marsh (2002).

\section{ Archival IUE Spectral Data }

All the spectral data were obtained from the Multimission Archive at Space Telescope (MAST) IUE archive are in a high activity state, very near or at outburst.  We restricted 
our selection to those systems with SWP spectra, with resolution of 5\AA\ and a spectral 
range of 1170\AA\ to 2000\AA.  All spectra were taken through the large aperture at 
low dispersion.  The Massa \& Fitzpatrick (2000) flux calibration-correction algorithm was 
applied to all the IUE data used; please see their paper for a description 
of the corrections it makes to the data.  When more than one spectrum with 
adequate signal-to-noise ratio was available, the two best spectra were analyzed.  An 
observing log of the observations is given in Table 1, where we list: (1) 
the system name, (2) data ID, (3) date of observation, 
(4) Universal Time of observation, (5) exposure time, and (6) activity 
state.  Transition refers to an intermediate state between outburst and quiescence, 
and an asterix in the last column indicates that an outburst state was determined 
without a ground-based optical light curve at the time of the IUE observation.

The activity state of the spectra was determined by examining the AAVSO
light curves for each system as well as the flux level of the IUE spectrum
that typically made it obvious the spectrum was obtained in outburst.
Unfortunately, we were unable to cross-reference this AAVSO data with
data from VSNET during the preparation of this paper due to their archives
being off-line.  In the case of those systems not covered by the AAVSO, their activity state was assessed based upon either mean photometric magnitudes taken from the Ritter \& Kolb(2003) catalogue or from IUE Fine Error Sensor (FES)  measurements at the
time of the IUE observation. The FES counts, when available, were
converted to optical magnitudes to help ascertain the brightness state at
the time of the IUE observation, and used as consistency checks on our
model fitting since the model-predicted optical magnitude from a
best-fitting model should always be fainter than observed optical
magnitude of the system. This is becuase of the contribution in the optical of the secondary. In addition, the presence of P-Cygni profiles, 
absorption lines, and comparison with spectral data and flux levels for the systems 
during other activity states was used to ascertain the state of the system.

The reddening of the systems was determined based upon all estimates listed in the literature. The three principal sources of reddening were
the compilations of la Dous (1991), Verbunt (1987) and Bruch \& Engel
(1999). If there was a range of values found, then the value we assume in
our analyses follows the range shown in Column (7) of Table 2, and the
ranges and assumed values are separated by a semicolon.  If only one value
was listed in the literature, then we adopted that value, and if there was
none listed, we took the reddening to be zero. The spectra were then 
de-reddened with the IUERDAF routine UNRED.

\section{Synthetic Spectral Fitting Procedure}
\subsection{Distance Constraints on the Fitting}

As a constraint on the goodness of a particular model fit, 
we adopted a distance from either a trigonometric parallax, the Warner (1995) and Harrison et al. (2004) relations or from the scale factor of the best-fitting WD model to HST spectra of dwarf novae during quiescence. The adopted distance was used as a basis for comparison with distances yielded by the accretion disk model fitting. The most direct means of knowing the distance is to have 
trigonometric parallaxes such as those measured by Thorstensen (2003) for 14 
dwarf novae. But dwarf novae and nova-like variables also offer the 
advantage of estimating distances from correlations between their absolute 
magnitudes at maximum brightness and their orbital periods. Distance estimates 
were obtained from either the absolute magnitude at outburst $\rm{M}_{V(max)}$ 
versus orbital period $P_{orb}$ relation of Warner (1995) or from a more recent 
relationship by Harrison et al. (2004) based upon their recent HST FGS parallaxes 
of dwarf novae. The Warner (1995) relation is
\begin{equation}
\rm{M}_{V(max)}=5.74-0.259\rm{P}_{orb} ({\rm hr})
\end{equation}
and the Harrison et al.(2004) relation is 
\begin{equation}
\rm{M}_{V(max)}=5.92-0.383\rm{P}_{orb} ({\rm hr})
\end{equation}
At the outset of the modeling process, we applied both of these 
relationships to each system.  Then, we conducted an exhaustive search of 
the literature for previous distance estimates.  If the literature search 
revealed other estimates, then we adopted some reasonable mean based upon 
the different methods used to obtain each distance estimate and the 
distance computed from the two calibrated $\rm{M}_{V(max)}-P_{orb}$ 
relations. If no other distance estimates existed, then we simply adopted 
a reasonable mean of the two $\rm{M}_{V(max)}-P_{orb}$ relations.  If a 
trigonometric parallax was available, then we adopted it for the distance.  
In the case of two systems (U Gem and YZ Cnc), there was more than one 
parallax.  For U Gem, all were identical.  For YZ Cnc, they were slightly 
different, so we took a mean of the parallax values.  In the case of SU 
UMa, the parallax obtained was unreliable, so we combined it in a mean 
with other distances estimates and the $\rm{M}_{V(max)}-P_{orb}$ relations.
The adopted distances used as constraints in the synthetic spectral fitting procedure
are shown in Table 2.

\subsection{Synthetic Spectral Fitting}

We adopted model accretion disks from the optically thick disk model grid 
of Wade \& Hubeny (1998).  In these accretion disk models, the innermost disk radius, 
R$_{in}$, is fixed at a fractional white dwarf radius of $x = R_{in}/R_{wd} = 1.05$. 
The outermost disk radius, R$_{out}$, was chosen so that T$_{eff}(R_{out})$ is near 
10,000K since disk annuli beyond this point, which are cooler zones with larger radii, 
would provide only a very small contribution to the mid and far UV disk flux, 
particularly the SWP FUV bandpass. The mass transfer rate is assumed to be the same for 
all radii.  Thus, the run of disk temperature with radius is taken to be:

\begin{equation}
T_{eff}(r)= T_{s}x^{-3/4} (1 - x^{-1/2})^{1/4}
\end{equation}

where  $x = r/R_{wd}$
and $\sigma T_{s}^{4} =  3 G M_{wd}\dot{M}/8\pi R_{wd}^{3}$

Limb darkening of the disk is fully taken into account in the manner described by 
Diaz et al. (1996) involving the Eddington-Barbier relation, the increase of kinetic 
temperature with depth in the disk, and the wavelength and temperature dependence 
of the Planck function. The boundary layer contribution to the model flux is not 
included. However, the boundary layer is expected to contribute primarily in the 
extreme ultraviolet below the Lyman limit.  

After masking emission lines, artifacts and poor quality data in the spectra, we determined separately for each system, the best-fitting accretion disk model 
using IUEFIT, a $\chi^{2}$ minimization routine. In virtually every dwarf nova system in our IUE sample, the accretion disk by itself provided very good agreement with the outburst spectra. However, if the accretion disk fit to the outburst spectrum was not entirely satisfactory over the entire wavelength range despite the disk providing nearly 100\% of the flux, then an attempt was made to slightly improve the fit by adding the flux contribution of a hot white dwarf. For this minority of cases,
WD Model spectra with solar abundances were created for high gravity stellar 
atmospheres using TLUSTY (Hubeny 1988) and SYNSPEC (Hubeny \& Lanz 1995).  
We took a range of gravities in the fitting from Log $g = 7.0 - 9.0$ in 
steps of 0.5.  For the white dwarf radii, we use the mass-radius relation from 
the evolutionary model grid of Matt Wood (private communication). These models have realistic non-degenerate H-rich envelopes, use the Lamb and Van Horn equation of state
and include the effect of non-zero temperature on the WD radius.

We took the best-fitting accretion disk model and combined it with a hot WD model, 
using a $\chi^{2}$ minimization routine called DISKFIT. The best-fitting composite accretion disk plus white dwarf  
was determined based upon the minimum $\chi^{2}$ value achieved 
and consistency of the scale factor-derived distance with the adopted 
distance for each system.  The scale factor, $S$, normalized to a kiloparsec 
and solar radius, can be related to the white dwarf radius R through:  
$F_{\lambda(obs)} = S H_{\lambda(model)}$, where $S=4\pi R^2 d^{-2}$, and $d$ is the 
distance to the source.

An illustration of the accuracy we can obtain in a formal error analysis with 
confidence contours are given in Winter \& Sion (2003). In that paper we 
presented accretion rates of EM Cygni, CZ Ori and WW Ceti in quiescence, 
in which $1\sigma, 2 \sigma$ and $3 \sigma$ confidence contours are given. 
In Table 2, we list the final composite fitting results for all systems, 
by column, as follows: (1) A reference number, (2) system name, (3) 
system type (and sub-type, if applicable), (4) $\rm{P}_{orb}$, 
(5) E(B-V), (6) adopted distance, (7) inclination, (8) mass accretion 
rate $\dot{\rm{M}}$, (9) percent flux contribution of the WD, and (10) 
percent flux contribution of the disk.  

The best fitting accretion disk model for each system is shown in the accompanying Figures 1-8, listed in order of increasing orbital period as in Table 2.  The thick solid line is the best fitting accretion disk.  The accretion disk is the overwhelmingly dominant contributor in the FUV. If in rare cases a hot white dwarf flux contribution was added in an attempt to slightly improve the fits, then the thick solid line represents their combination, the dashed curve represents the accretion disk alone, and the dotted curve represents the white dwarf.  Also included are a reference number and name given in Table 2 to clearly identify each system.  To discuss a particular system, the figures are labelled Figure N.nn, where N is the figure number and nn is the reference number of a particular dwarf nova.  

There are several distances listed in column (6) of Table 2 that are
followed by either a $\pi$, or a $\pi$ in parentheses.  If the distance is
followed by a $\pi$, that distance is taken from a parallax (or parallaxes
in the case of U Gem).  If the distance is followed by a $\pi$ in
parentheses, then it is an adopted distance that incorporates a parallax
(see the explanation in section 3.1).  All parallaxes are taken from
either Harrison et al. (2004), or Thorstensen (2003). A colon in Column 7 denotes an  uncertain result.

Although the eclipsing dwarf novae OY Car and Z Cha are part of the IUE archival
sample, these systems are not included in the statistics of this study and their
derived accretion rates cannot be taken seriously. This is because the accretion
disk in these systems is self-eclipsed (i.e. the outer disk obscures the inner disk). Hence, any derived parameters (for example, the accretion rate, percentage of disk flux contribution) are systematically affected.  

In Table 2 it is seen that virtually all of the accretion rates derived from the 
FUV flux distribution with steady state, optically thick disk models fall within 
the range $1\times10^{-8}$ to $1\times10^{-9} \rm{M}_{\sun}yr^{-1}$ for the 
distances, white dwarf masses, orbital inclinations, and interstellar reddening 
values we have adopted. This is no great surprise although the range of accretion 
rates that we derive is somewhat broader than the range derived by Warner (1995) 
whose accretion rates were based upon the absolute magnitudes of the accretion 
disk corrected for inclination and limb darkening effects. It should be acknowledged however that some of the distances for the 44 dwarf novae in our study involved the use of the Warner (1995) relation. Hence, the rough agreement with the range of accretion rates found by Warner (1995) from his optical study is biased to some degree in favor of such an agreement. Moreover, as is well known, in all of the outburst spectra, almost without exception, \ion{N}{5} (1238\AA, 1242\AA), \ion{C}{4} (1548\AA, 1550\AA) and 
\ion{Si}{4} (1393\AA, 1402\AA), the hallmark resonance doublets are seen in 
prominent blue-shifted absorption and originate in the CV wind. In many of the 
hotter disk models, one sees synthetic Si IV absorption but the disk temperatures 
are not high enough for significant non-wind \ion{C}{4},  \ion{N}{5} or 
\ion{O}{5}. The strongest metallic absorption seen in the spectra, aside from 
the wind resonance doublets, is due to a blend of multiplet 2 of \ion{Si}{3} 
(1300\AA), \ion{Si}{2} and \ion{O}{1} (1302\AA). In virtually all of the 
model accretion disk fits, the observed absorption centered at 1300\AA\ 
is reasonably well represented by the synthetic disk \ion{Si}{3} absorption at 
the solar abundance of silicon. A difficulty of fitting the width of the Lyman 
$\alpha$ profiles in most of the outburst spectra is the presence of \ion{N}{5} 
which complicates discerning the longward wing of Lyman $\alpha$. Fortunately, 
the shortward wing is usually quite apparent and one can assume, as a first 
approximation, mirror symmetry between the two wings.  Nonetheless, in some cases 
the \ion{N}{5} looks suspiciously like a pseudo-absorption feature created by the 
geocoronal emission reversal seen in the core of Lyman $\alpha$.  

Several individual spectra are noteworthy for reasons other than just the goodness 
of their fit with steady state, optically thick accretion disk energy distributions. 
In Figure 1.02, the outburst spectrum of the SU UMa system SW UMa along 
with the best-fitting accretion disk model. There are profiles (e.g. \ion{C}{4} 
1550\AA, \ion{Si}{2} 1260\AA) which appear to have emission flanking absorption on both the shortward and longward sides of the absorption. Similar profile shapes were seen 
during the early stages of the July 2001 superoutburst of WZ Sge (Sion et al.2003). The emission seen near \ion{C}{4} could well be due to the \ion{Si}{2} (1526\AA, 1533\AA) doublet, which would match the same structure seen at \ion{Si}{2} (1260\AA). If this 
is the case, then \ion{C}{4} is barely seen and may be very weak relative to nitrogen.

In Figures 1.04, 3.16, 4.18, 6.33, 7.35, 7.39, 7.40, 8.41, 8.42, 8.43, and 8.44, the outburst spectra of T Leo, WX Hyi, YZ Cnc, RX And, AH Her, Z Cam, EM Cyg, RU Peg, SY Cnc, DX And, and BV Cen, respectively, reveal a deviation from a steady state flux distribution 
longward of 1700\AA. In every case, the disk model flux is lower than the observed flux.
The deviation tends to be more pronounced among Z Cam-type systems.
A anonymous referee suggests this may be due to the secondary star or possibly a shortfall of flux due to the temperature of the outermost disk boundary being set at 10,000K.
In our sample, except for T leo, this deviation is seen primarily in systems above the period gap. 

As stated earlier, one difficulty of fitting the width of 
the Lyman $\alpha$ profiles in this and other outburst spectra is the presence 
of \ion{N}{5} (1238\AA, 1242\AA) which is some cases may actually be a pseudo-feature 
created by the geocoronal emission reversal seen in the core of Lyman $\alpha$. 
In Figure 2.09, TV Crv has an unusually strong \ion{Si}{4} (1393\AA, 1402\AA) 
absorption suggesting possible suprasolar Si abundance or a higher \ion{Si}{4} ionization 
fraction in the outflowing wind material.

\section{Discussion}

Our analyses represent the first detailed look at all of the dwarf novae observed with the IUE SWP in outburst in a systematic way. We have applied a uniform analysis to the 
largest spectroscopic sample of FUV spectra of dwarf novae, obtained during the same 
brightness state with the same telescope, and the same instrumental setup, taking 
into the account the best available distance information. Accordingly, this large 
body of accretion rates will serve as a basis of comparison with accretion rates derived 
by other independent methods in other wavelength regions. Moreover, the FUV-derived 
accretion rates of individual systems can be compared with FUV-derived accretion rates 
of other systems in the same dwarf nova subclass or across subclasses. A number of 
implications and conclusions arise from this analysis.

Overall, the steady state, optically thick disk model fits to the outburst 
spectra shown in Figures 1 through 8 are quite acceptable and in some cases excellent. 
All of the accretion rates derived from the FUV fall within the range $1\times10^{-8}$ to 
$1\times10^{-9} \rm{M}_{\sun}yr^{-1}$ for the distances, white dwarf masses, 
orbital inclinations, and reddening values we have adopted.  Therefore, the main 
uncertainties in the derived accretion rates, or in constraining the outburst accretion 
rates, arise from the uncertainties in the published values of the white dwarf 
masses, the orbital inclinations and the reliability of the distances.

On the other hand, there are several systems with outburst spectra in the IUE archive 
that merit attention because they differ significantly from the FUV energy distribution 
of steady state accretion disks. Aside from the eclipsing systems Z Cha and OY Car, which 
reside below the period gap and are seen nearly edge-on with absorbing curtains, all of 
the systems that depart from a steady state disk distribution do so longward of 1600\AA 
and all of these systems lie above the period gap except for T Leo. The U Gem-type and Z Cam-type systems are WX Hyi, YZ Cnc, RX And, AH Her, Z Cam, EM Cyg, RU Peg, SY Cnc, DX And, and BV Cen.

If one takes our FUV-derived accretion rates in outburst at face value, then in principle
they can be used in combination with accretion rates derived for dwarf nova quiescence,
and the timescales of outburst and quiescence to estimate the time-averaged accretion 
rate of the system during its lifetime as a dwarf nova. That is, when average accretion 
rates are known during quiescence and during outburst, then a true time-averaged 
accretion rate or mean accretion rate, $<\dot{\rm{M}}>$, may be derived. If 
$\dot{\rm{M}}_{ob}$ is an average accretion rate in outburst, $\dot{\rm{M}}_q$ is an 
average accretion rate in quiescence, $t_{ob}$ is the average duration of outburst 
and $t_{q}$ is the average duration of quiescence, then for a dwarf nova duty cycle
\begin{equation}
t_{tot} = t_{ob} + t_{q}
\end{equation}
the "duty-cycle" averaged $\dot{M}$ is given by
\begin{equation}
<\dot{\rm{M}}>=[\dot{\rm{M}}_{ob}t_{ob} + \dot{\rm{M}}_qt_q] t_{tot}^{-1}
\end{equation}
Now it appears that all dwarf novae both above and below the gap have 
$\dot{\rm{M}}=10^{-8}$ to $10^{-9} \rm{M}_{\sun}yr^{-1}$ in outburst. Suppose we 
assume the outburst times and quiescence times as well as the WD masses of all 
dwarf novae are the same. Then the only possible causes for a difference in 
accretion heating of a WD in a DN above the gap compared with a WD below the gap 
are: (1) a difference in WD ages due to evolution age; (2) a different amount of 
disk accretion during the outburst (i.e. more mass from the disk accreted above the gap 
versus the mass fraction of the disk accreted below the gap) or; (3) a significantly 
higher accretion rate during quiescence for DN above the gap than below the gap. 
Looking at the above equation, if $\dot{\rm{M}}_q$ is sufficiently high and the 
quiescence long enough (as in many long period dwarf novae), then the second term 
in equation (4) can compete and possibly overwhelm the first term if the outburst 
duration is short enough, if $\dot{\rm{M}}_{ob}$ is lower, say, less than 
$10^{-9} \rm{M}_{\sun}yr^{-1}$, or if the accretion rate during quiescence is 
higher than a nominal $1\times10^{-11} \rm{M}_{\sun}yr^{-1}$ that we assume in this paper.

While the above time-averaged accretion rate are gotten directly from model-derived 
accretion rates, the time-averaged accretion rates derived for selected dwarf novae 
by Patterson (1984) were obtained for the most part from a mean optical brightness level
in AAVSO light curves on the assumption that the accretion disk is the sole source of 
the system light. By converting the mean magnitude to a flux and combining it with an 
estimated distance, a luminosity was obtained. The accretion rate providing that luminosity was derived for a given assumed or known white dwarf mass. 

It is of interest to carry out a quantitative comparison of the two approaches to 
estimating $<\dot{\rm{M}}>$. For illustration, we take sample values for the 
outburst timescale and the recurrence timescale of a typical dwarf nova. We assume 
for simplicity that the recurrence time of dwarf nova outbursts is constant at, 
say 21 days, and the outburst timescale is constant at 7 days. For these parameters, 
if the outburst accretion rate is $10^{-8} \rm{M}_{\sun}yr^{-1}$ then, by (1), 
$<\dot{\rm{M}}>=2.5\times10^{-9} \rm{M}_{\sun} yr^{-1}$. If the outburst accretion 
rate is $10^{-9} \rm{M}_{\sun} yr^{-1}$ and the quiescent rate is an assumed to 
be $1\times10^{-11} \rm{M}_{\sun}yr^{-1}$, then 
$<\dot{\rm{M}}>=2.5\times10^{-10} \rm{M}_{\sun}yr^{-1}$. 

Suppose we take average values of $t_{ob}$ and $t_q$ for two well-studied systems 
such as U Gem and SS Cygni. For U Gem, from Table 2, we take $\dot{\rm{M}}_{ob}$ 
to be $2\times10^{-8} \rm{M}_{\sun}yr^{-1}$, $t_{ob}$ of 14 days (wide outburst), 
$t_q$ of 100 days and $\dot{\rm{M}}_q$ of $1\times10^{-11} \rm{M}_{\sun}yr^{-1}$. 
Hence for these parameters U Gem would have  
$<\dot{\rm{M}}>\simeq2.5\times10^{-9} \rm{M}_{\sun}yr^{-1}$. Considering SS Cygni 
with the same quiescence accretion rate, $t_{ob}$ of 14 days, $t_q$ of 40 days 
and $\dot{\rm{M}}_{ob}$ of $3\times10^{-9} \rm{M}_{\sun}yr^{-1}$ from Table 2, 
we find that $<\dot{\rm{M}}>\sim7\times10^{-10} \rm{M}_{\sun}yr^{-1}$.

Patterson (1984) estimated $<\dot{\rm{M}}>\simeq3\times10^{-10} \rm{M}_{\sun}yr^{-1}$ 
for U Gem and $<\dot{\rm{M}}>\sim5\times10^{-10} \rm{M}_{\sun}yr^{-1}$ for SS Cygni. 
We note that the distances adopted by Patterson (1984), 78 pc and 95 pc for U Gem 
and SS Cygni respectively, are closer than the FGS parallax distances we used for the 
two systems. Comparing Patterson's values with the time-averaged accretion rates 
estimated using Table 2, our value of $<\dot{\rm{M}}>$ for U Gem is 10 times higher 
than estimated in Patterson (1984) while for SS Cygni, our $<\dot{\rm{M}}>$ is 1.4 
times higher than the value estimated by Patterson (1984). Until the accretion rates 
during dwarf nova quiescence are more accurately known, it is premature to compute 
the time-averaged accretion rates in this direct manner. Moreover, individual accretion 
rates will not be secure for dwarf novae until there is further progress in determining
better white dwarf masses, orbital inclinations and distances. 

We are grateful to an anonymous referee for helpful comments.  This work was supported 
by NSF grant AST05-07514, NASA grant NNG04GE78G and by 
summer undergraduate research support from the Delaware Space Grant Consortium.  
Some or all of the data presented in this paper were obtained from the Multimission 
Archive at the Space Telescope Science Institute (MAST). STScI is operated by the 
Association of Universities for Research in Astronomy, Inc., under NASA contract 
NAS5-26555. Support for MAST for non-HST data is provided by the NASA Office of Space 
Science via grant NAG5-7584 and by other grants and contracts.

\clearpage

\begin{deluxetable}{lccccc}
\tabletypesize{\tiny}
\tablewidth{0.0pt}
\tablenum{1}
\tablecaption{\sc{Observing Log}}
\tablecolumns{7}
\tablehead{
        \colhead{System} &
        \colhead{Data ID} &
        \colhead{Observation Date} &
        \colhead{Observation Time} &
        \colhead{Exposure Time} &
        \colhead{Outburst-Quiescence Cycle} \\
   \colhead{\null} &
   \colhead{\null} &
   \colhead{(yyyy-mm-dd)} &
   \colhead{(UT)} &
  \colhead{(s)} &
   \colhead{\null} \\
\colhead{(1)} &
\colhead{(2)} &
\colhead{(3)} &
\colhead{(4)} &
\colhead{(5)} &
\colhead{(6)} \\

}

\startdata

   WZ Sge & SWP03507 & 1978-12-01 & 19:21:00 & 24   & Peak Outburst \\
   SW UMa & SWP27871 & 1986-03-08 & 04:32:00 & 210  & Outburst\\
   WX Cet & SWP36511 & 1989-06-15 & 10:41:00 & 3300 & Late Decline \\
    T Leo & SWP33642 & 1988-05-25 & 20:53:00 & 300  & Outburst \\
V1159 Ori & SWP56889 & 1996-03-03 & 17:01:00 & 5400 & Rise to Outburst \\
 V436 Cen & SWP54246 & 1995-03-28 & 10:09:00 & 2400 & Outburst $\ast$ \\
   BC UMa & SWP50668 & 1994-05-01 & 21:09:00 & 3000 & Peak Outburst \\
   EK TrA & SWP09705 & 1980-08-05 & 22:20:00 & 1140 & Outburst $\ast$ \\
   TV Crv & SWP41842 & 1991-06-14 & 14:25:00 & 1800 & Outburst $\ast$ \\
   OY Car & SWP25838 & 1985-05-02 & 17:34:00 & 720  & Rise to Outburst \\
   VY Aqr & SWP21720 & 1983-12-08 & 03:22:00 & 600  & Outburst \\
   ER UMa & SWP40947 & 1991-02-27 & 22:20:00 & 4500 & Outburst $\ast$ \\
   IR Gem & SWP38524 & 1990-04-05 & 02:03:00 & 1068 & Outburst \\
   AY Lyr & SWP09342 & 1980-06-22 & 06:46:00 & 3600 & Late Outburst \\
   VZ Pyx & SWP44129 & 1992-03-07 & 05:36:00 & 2400 & Outburst $\ast$ \\
   VW Hyi & SWP40045 & 1990-11-04 & 10:45:00 & 150  & Peak Outburst \\
    Z Cha & SWP30670 & 1987-03-31 & 17:03:00 & 600  & Peak Outburst \\
   WX Hyi & SWP23952 & 1984-09-13 & 19:22:00 & 3600 & Outburst $\ast$ \\
   SU UMa & SWP35261 & 1989-01-05 & 08:58:00 & 1800 & Peak Outburst \\
   YZ Cnc & SWP15560 & 1981-11-24 & 10:07:00 & 540  & Outburst \\
   TU Men & SWP10665 & 1980-11-23 & 17:58:00 & 1800 & Outburst $\ast$ \\
   AB Dra & SWP07413 & 1979-12-15 & 18:51:00 & 4500 & Rise to Outburst \\
   CY Lyr & SWP21030 & 1983-09-12 & 22:09:00 & 3600 & Outburst \\
   KT Per & SWP07382 & 1979-12-13 & 01:19:00 & 2700 & Early Decline \\
   AR And & SWP15445 & 1981-11-07 & 12:34:00 & 2100 & Outburst \\
   CN Ori & SWP15950 & 1982-01-04 & 08:36:00 & 2400 & Outburst \\
    X Leo & SWP15985 & 1982-01-06 & 14:03:00 & 6240 & Outburst \\
   VW Vul & SWP18875 & 1982-12-23 & 19:08:00 & 7200 & Transition $\ast$\\
   UZ Ser & SWP15078 & 1981-09-22 & 18:41:00 & 2400 & Outburst \\
    U Gem & SWP10327 & 1980-10-10 & 13:28:00 & 353  & Peak Outburst \\
   TW Vir & SWP10951 & 1981-01-03 & 06:47:00 & 3600 & Outburst $\ast$ \\
   SS Aur & SWP18610 & 1982-11-20 & 17:33:00 & 1560 & Rise to Outburst \\
   HX Peg & SWP37459 & 1989-10-25 & 19:31:00 & 3600 & Outburst $\ast$ \\
   RX And & SWP17678 & 1982-08-13 & 22:59:00 & 600  & Peak Outburst \\
   CZ Ori & SWP16039 & 1982-01-14 & 03:18:00 & 2700 & Outburst \\
   AH Her & SWP17662 & 1982-08-11 & 22:12:00 & 1200 & Outburst \\
   TZ Per & SWP17643 & 1982-08-09 & 19:37:00 & 2580 & Outburst \\
   TT Crt & SWP47805 & 1993-06-04 & 19:54:00 & 3300 & Outburst \\
   SS Cyg & SWP47746 & 1993-05-27 & 07:57:00 & 180  & Outburst \\
    Z Cam & SWP21721 & 1983-12-08 & 05:41:00 & 240  & Early Decline \\
   EM Cyg & SWP07297 & 1979-12-02 & 22:45:00 & 1500 & Outburst \\
   RU Peg & SWP15079 & 1981-09-22 & 20:43:00 & 840  & Decline \\
   SY Cnc & SWP07313 & 1979-12-05 & 00:40:00 & 1200 & Mid Decline \\
   DX And & SWP37687 & 1989-11-26 & 15:25:00 & 2700 & Rise to Outburst \\
   WW Cet & SWP10664 & 1980-11-23 & 16:21:00 & 1800 & Transition $\ast$\\
   BV Cen & SWP24867 & 1985-01-08 & 20:48:00 & 3600 & Transition $\ast$\\
\enddata
\caption{Sample stars and associated SWP spectra listed by increasing orbital period.  
         An asterix in the final column denotes that the cycle state was determined in the 
         absence of an optical light curve as described in \S{2}.}
\end{deluxetable}

\begin{deluxetable}{clclccclcc}
\tabletypesize{\tiny}
\tablewidth{0.0pt}
\tablenum{2}
\tablecaption{\sc{System Parameters}} 
\tablecolumns{10}
\tablehead{
        \colhead{Reference} &
        \colhead{System} &
        \colhead{Sub Type} &
        \colhead{Orbital period} &
        \colhead{Reddening} &
        \colhead{Distance} &
        \colhead{$i$} &
        \colhead{\.{M}} &
        \colhead{WD \%} &
        \colhead{Disk \%} \\
   \colhead{Number} &
   \colhead{\null} &
   \colhead{\null} &
   \colhead{(days)} &
   \colhead{\null} &
   \colhead{(pc)} &
   \colhead{($\degr$)} &
   \colhead{($\rm{M}_{\sun}yr^{-1}$)} &
   \colhead{\null} &
   \colhead{\null} \\
\colhead{(1)} &
\colhead{(2)} &
\colhead{(3)} &
\colhead{(4)} &
\colhead{(5)} &
\colhead{(6)} &
\colhead{(7)} &
\colhead{(8)} &
\colhead{(9)} &
\colhead{(10)} \\
}
\startdata

01&  WZ Sge   &SU-WZ      &0.056687846  &0.00           &43.3$\pi$&      78&$2.5\times10^{-8}$     &$\leq0.5$&$\geq$99.5    \\
02&  SW Uma   &SU(DQ?)    &0.056815     &0.00           &110      &      60&$3.2\times10^{-9}$     &0        &100           \\
03&  WX Cet   &SU-WZ      &0.05827      &0.00 ?         &187      &      41&$1.0\times10^{-9}$     &0        &100           \\
04&  T Leo    &SU         &0.05882      &0.00           &101$\pi$ &      60&$1.0\times10^{-8}$     &0        &100           \\
05&  V1159 Ori&SU-ER      &0.06217801   &0.00 ?         &213      &      60&$1.0\times10^{-9}$  :  &0        &100           \\
06&  V436 Cen &SU         &0.062501     &0.05; 0.00     &320      &      41&$2.2\times10^{-9}$     &1        &99            \\
07&  BC Uma   &SU-WZ      &0.06261      &0.00 ?         &285      &      75&$3.2\times10^{-9}$     &0        &100           \\
08&  EK TrA   &SU         &0.06288      &0.05;0.00      &200      &      18&$9.0\times10^{-10}$ :  &$\leq$1  &$\geq$99      \\
09&  TV Crv   &SU         &0.0629       &0.00           &268      &      60&$1.0\times10^{-9}$     &0        &100           \\
10&  OY Car   &SU         &0.062917     &0.00           &85       &      75&$6.0\times10^{-9}$     &$\leq$11 &$\geq$89      \\
11&   VY Aqr   &SU-WZ      &0.06309      &0.00           &97$\pi$  &      41&$1.0\times10^{-9}$     &0        &100           \\
12&  ER Uma   &SU-ER      &0.06366      &0.00 ?         &460      &      60&$9.0\times10^{-9}$  :  &$\leq$1  &$\geq$99      \\
13&  IR Gem   &SU         &0.0684       &0.00           &250      &      60&$1.0\times10^{-8}$     &0        &100           \\
14&  AY Lyr   &SU         &0.0737       &0.00           &380      &      41&$3.2\times10^{-9}$     &0        &100           \\
15&  VZ Pyx   &SU (IP?)   &0.07332      &0.00           &280      &      60&$2.8\times10^{-9}$     &5        &95            \\
16&  VW Hyi   &SU         &0.074271     &0.00           &65       &      60&$3.2\times10^{-9}$     &0        &100           \\
17&  Z Cha   &SU         &0.074499     &0.00           &97       &      81&$4.0\times10^{-9}$  :  &$\leq$19 &$\geq$81       \\
18&  WX Hyi   &SU         &0.0748125    &0.00           &280      &      60&$1.0\times10^{-9}$     &1        &99            \\
19&  SU Uma   &SU         &0.07635      &0.00           &270$\pi$ &      41&$7.0\times10^{-10}$    &2        &98             \\
20&  YZ Cnc   &SU         &0.0868       &0.00           &265$\pi$ &      41&$2.5\times10^{-9}$     &$\leq$2  &$\geq$98      \\
21&  TU Men   &SU         &0.1172       &0.08           &320      &      41&$1.0\times10^{-8}$     &0        &100           \\
22&  AB Dra   &ZC         &0.15198      &0.10           &400      &      81&$1.0\times10^{-9}$     &8        &92            \\
23&  CY Lyr   &UG         &0.1591       &0.18           &400      &      60&$1.0\times10^{-8}$     &0        &100           \\
24&  KT Per   &ZC         &0.16265777   &0.15-0.54; 0.2 &245      &      60&$2.2\times10^{-9}$     &$\leq$0.5&$\geq$99.5    \\
25&  AR And   &UG (SU?)   &0.16302      &0.02           &270      &      60&$1.0\times10^{-8}$  :  &       0 &100           \\
26&  CN Ori   &UG         &0.163199     &0.00           &295      &      60&$9.0\times10^{-9}$     &2        &98            \\
27&  X Leo    &UG         &0.1646       &0.00           &350      &      41&$3.2\times10^{-10}$    &12       &88            \\
28&  VW Vul   &ZC         &0.16870      &0.15           &650      &      41&$1.0\times10^{-8}$     &19       &81            \\
29&  UZ Ser   &UG (ZC?)   &0.173        &0.35           &300      &      18&$2.0\times10^{-8}$     &1        &99            \\
30&  WW Cet   &UG (ZC?)   &0.1758       &0.00           &190      &      60&$1.0\times10^{-9}$  :  &$\leq$1  &$\geq$99      \\
31&  U Gem    &UG         &0.17690619   &0.00-0.05; 0   &96$\pi$  &      75&$1.6\times10^{-8}$     &1        &99            \\
32&  TW Vir   &UG         &0.18267      &0.00           &500      &      60&$2.8\times10^{-9}$  :  &$\leq$1  &$\geq$99      \\
33&  SS Aur   &UG         &0.1828       &0.10           &201$\pi$ &      41&$7.0\times10^{-9}$  :  &$\leq$0.5&$\geq$99.5    \\
34&  HX Peg   &\nodata    &0.2008       &0.00           &550$(\pi)$&     41&$2.8\times10^{-9}$  :  &$\leq8.5$&$\geq$91.5    \\
35&  RX And   &ZC         &0.209893     &0-0.06; 0      &200      &      41&$2.0\times10^{-9}$     &1.5      &98.5          \\
36&  CZ Ori   &UG         &0.214667     &0.00           &260      &      18&$3.0\times10^{-10}$ :  &$\leq$14 &$\geq$86      \\
37&  AH Her   &ZC         &0.258116     &0.3            &660$\pi$ &      41&$9.0\times10^{-9}$  :  &$\leq$0.5&$\geq$99.5    \\
38&  TZ Per   &ZC         &0.262906     &0.27           &275      &      41&$6.0\times10^{-9}$     &5        &95            \\
39&  TT Crt   &UG         &0.2683522    &0.00           &500      &      60&$3.0\times10^{-9}$  :  &$\leq$3  &$\geq$97      \\
40&  SS Cyg   &UG         &0.27513      &0.04; 0        &166$\pi$ &      18&$3.2\times10^{-9}$     &$\leq$0.5&$\geq$99.5    \\
41&  Z Cam    &ZC         &0.2898406    &0.0-0.06; 0    &112$\pi$ &      60&$9.0\times10^{-10}$    &1        &99            \\
42&  EM Cyg   &ZC         &0.290909     &0.05; 0        &350      &      75&$7.0\times10^{-9}$     &$\leq$0.5&$\geq$99.5    \\
43&  RU Peg   &UG         &0.3746       &0.00           &282$\pi$ &      41&$1.0\times10^{-9}$     &0        &100           \\
44&  SY Cnc   &ZC         &0.38         &0.00           &450      &      18&$1.0\times10^{-8}$     &0        &100           \\
45&  DX And   &UG         &0.440502     &0.00           &500      &      60&$1.0\times10^{-8}$     &7        &93            \\
46&  BV Cen   &UG         &0.610108     &0.05-0.36; 0.1 &500      &      60&$3.2\times10^{-9}$     &$\leq$0.5&$\geq$99.5    \\
\enddata

\caption{Sample stars listed by increasing orbital period.  A reference number is given to each system to locate the associated 
fit in the accompanying figures.}
\end{deluxetable}

\clearpage

\begin{figure}
\epsscale{0.92}
\plotone{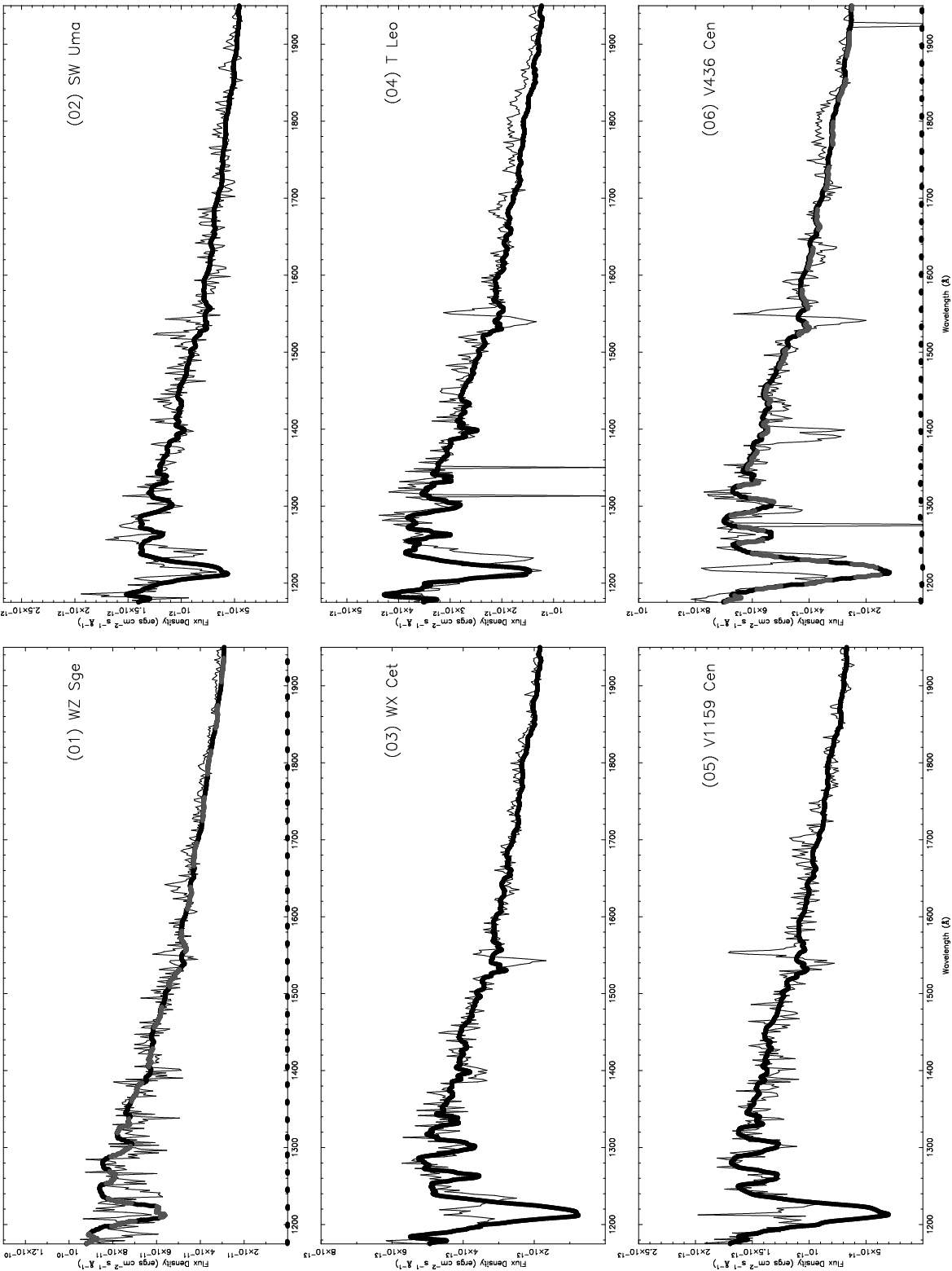}
\caption{Flux density versus wavelength plots for (01) WZ Sge, (02) SW UMa, (03) WX Cet,
(04) T Leo, (05) V1159 Cen, (06) V436 Cen. The best fitting accretion disk model is shown with the thick solid line. If in rare cases a hot WD model was added to the best-fit accretion disk to try to improve the fit, then the WD model flux is denoted by a dotted line, the accretion disk model flux alone by a dashed line and the combined flux by a thick solid line. All very sharp narrow absorption features seen extending down to zero flux level are due to a grid of fiducial reseaux marks located between the UV converter and the SEC Vidicon camera of IUE}.
\end{figure}
\clearpage

\begin{figure}
\epsscale{0.92}
\plotone{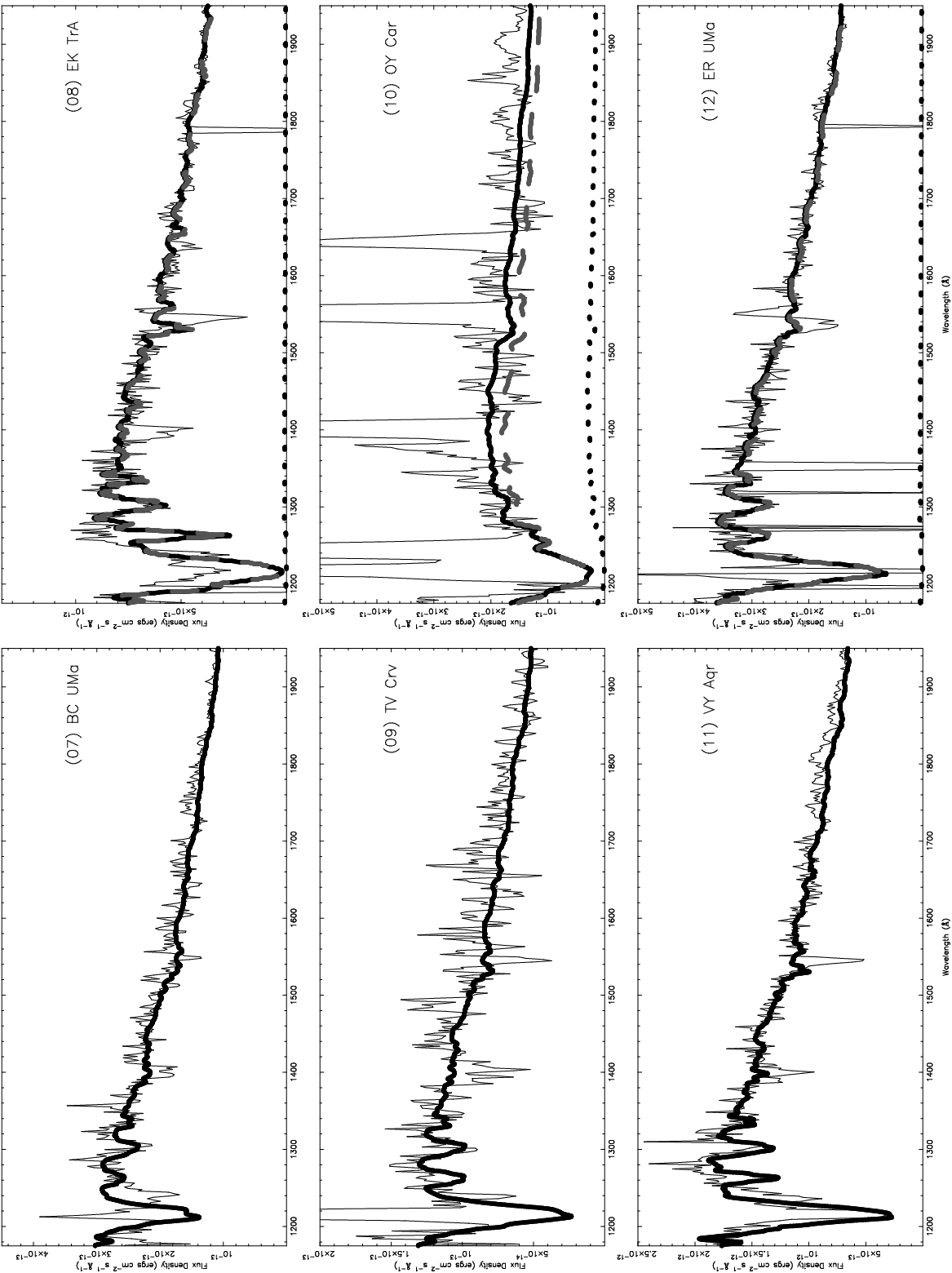}
\caption{Same as Figure 1 except for (07) BC UMa, (08) EK Tra, (09) TV Crv, (10) OY Car,
(11) VY Aqr, (12) ER UMa}
\end{figure}
\clearpage

\begin{figure}
\epsscale{0.92}
\plotone{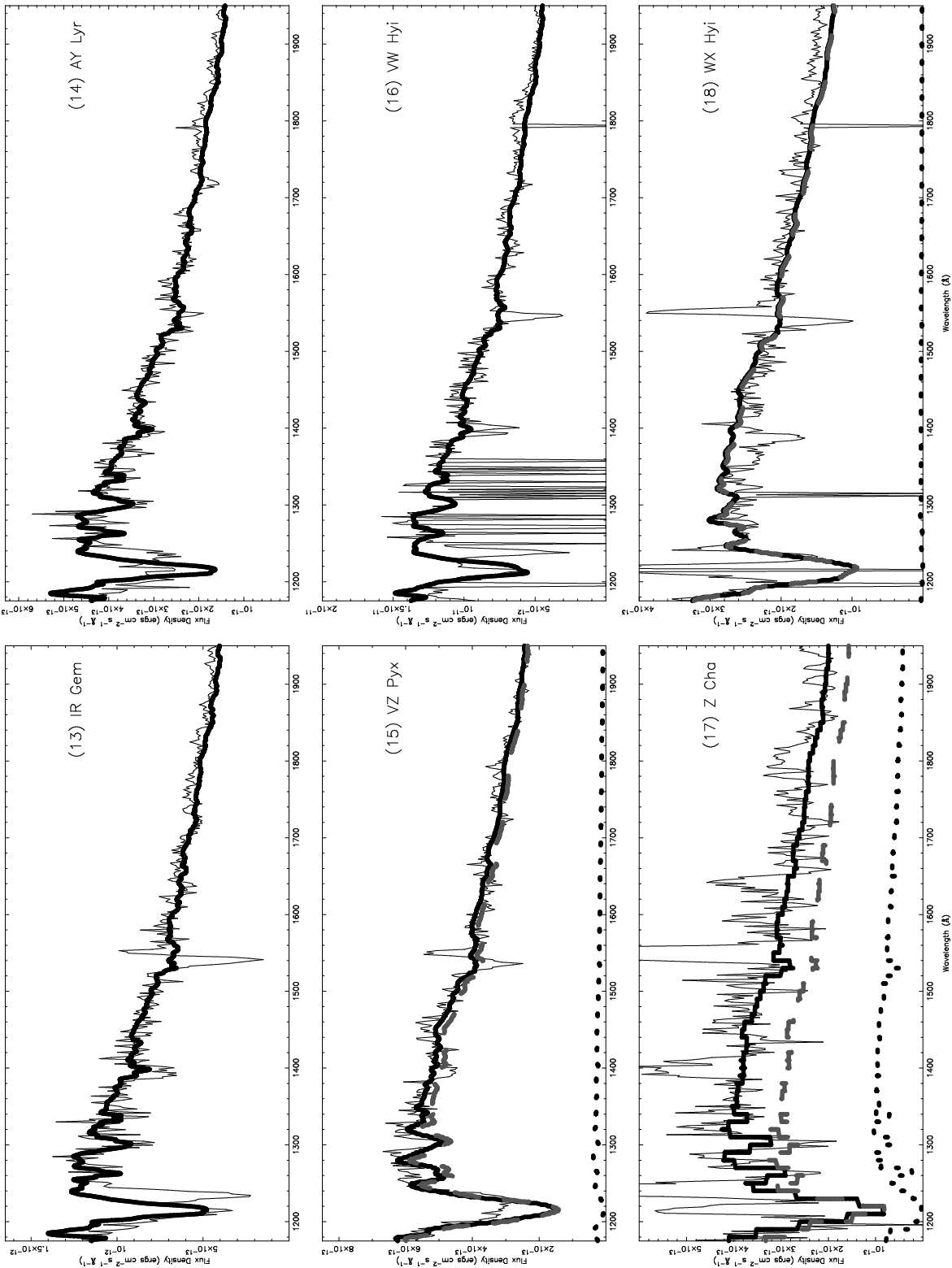}
\caption{Same as Figure 1 except for (13) IR Gem, (14) AY Lyr, (15) VZ Pyx, (16) VW Hyi,
(17) Z Cha, (18) WX Hyi}
\end{figure}
\clearpage

\begin{figure}
\epsscale{0.92}
\plotone{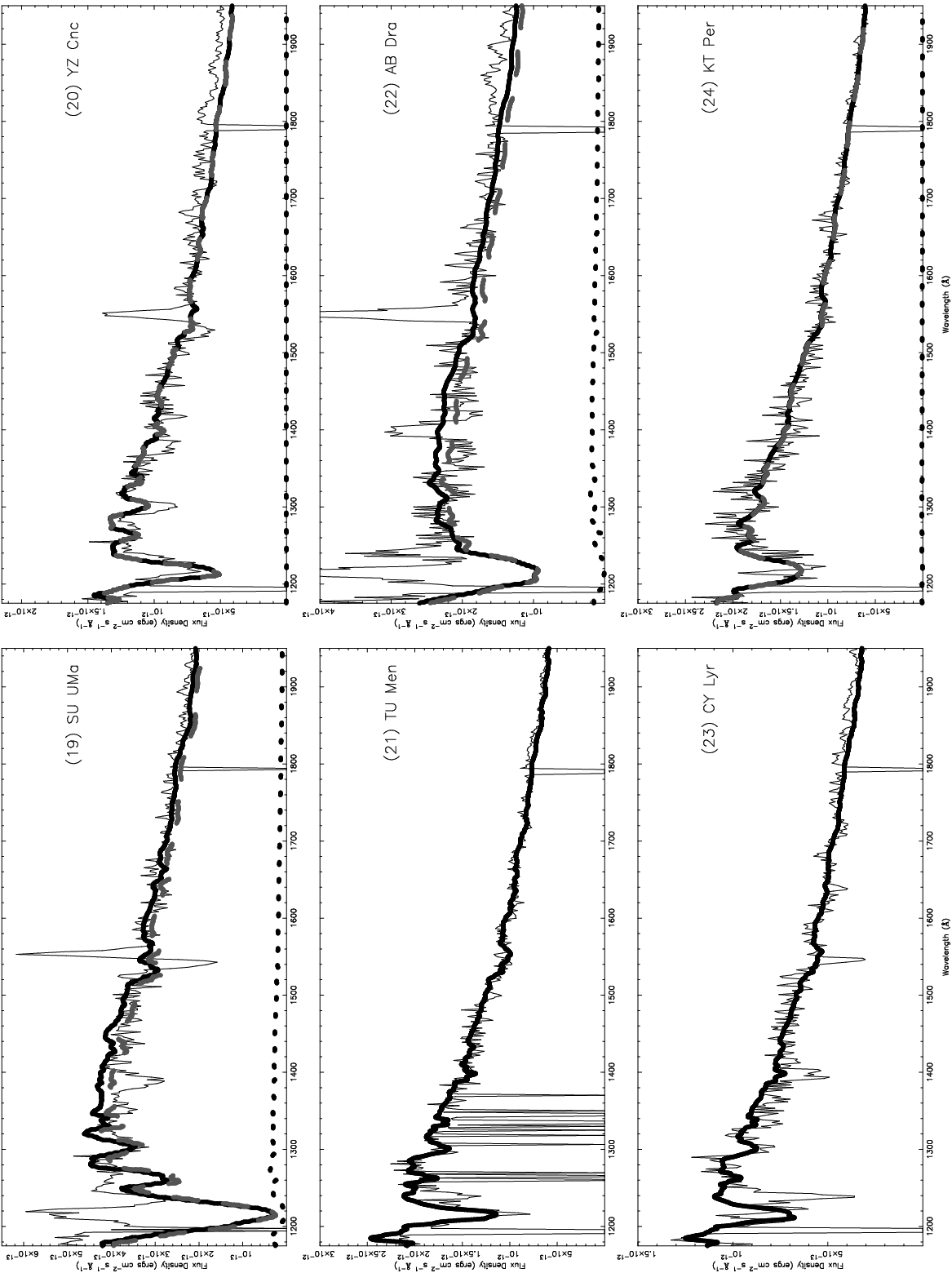}
\caption{Same as Figure 1 except for (19) SU UMa, (20) YZ Cnc, (21) TU Men, (22) AB Dra,
(23) CY Lyr, (24) KT Per}
\end{figure}
\clearpage

\begin{figure}
\epsscale{0.92}
\plotone{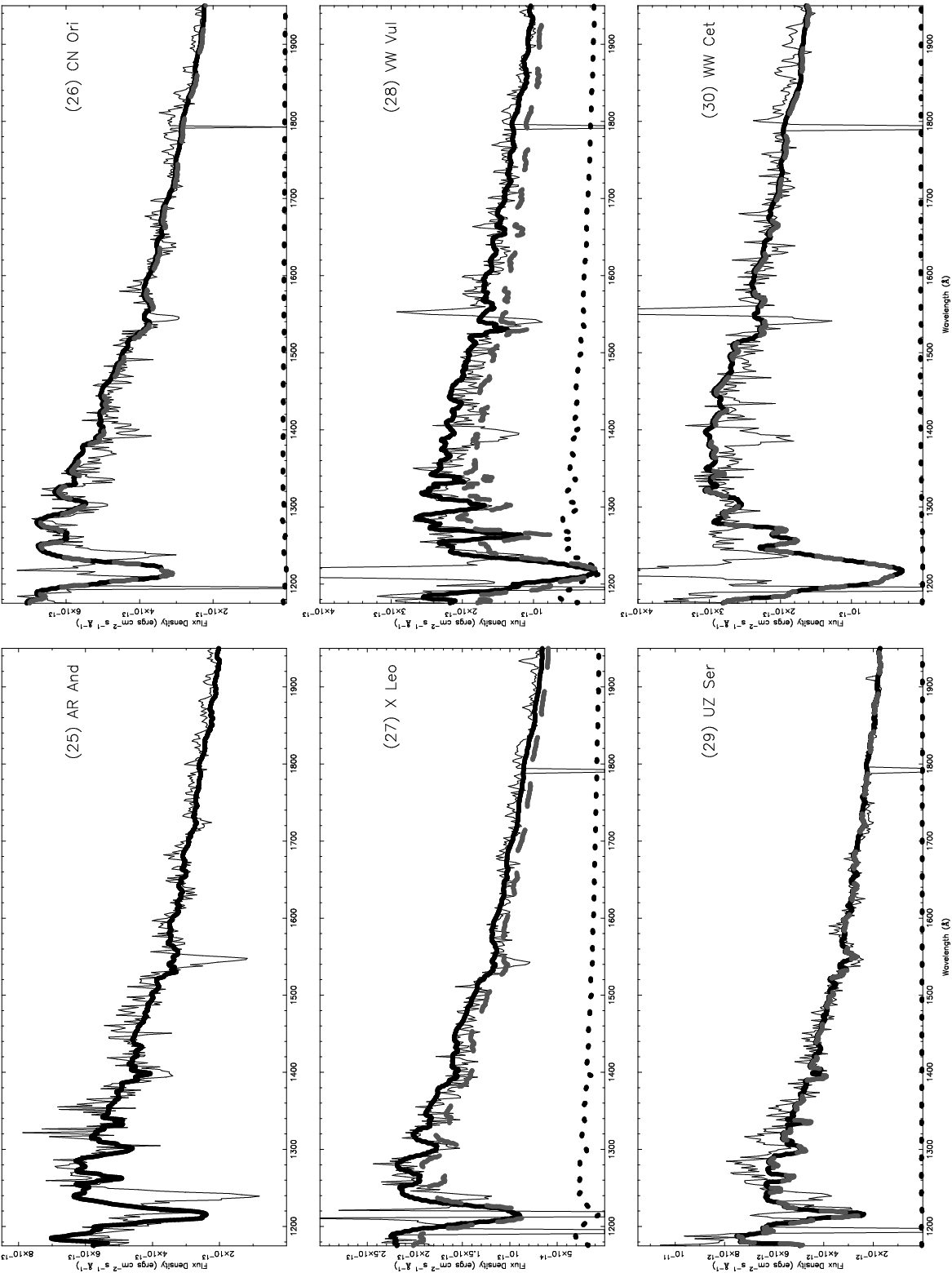}
\caption{Same as Figure 1 except for (25) AR And, (26) CN Ori, (27) X Leo, (28) VW Vul,
(29) UZ Ser, (30) WW Cet}
\end{figure}
\clearpage

\begin{figure}
\epsscale{0.92}
\plotone{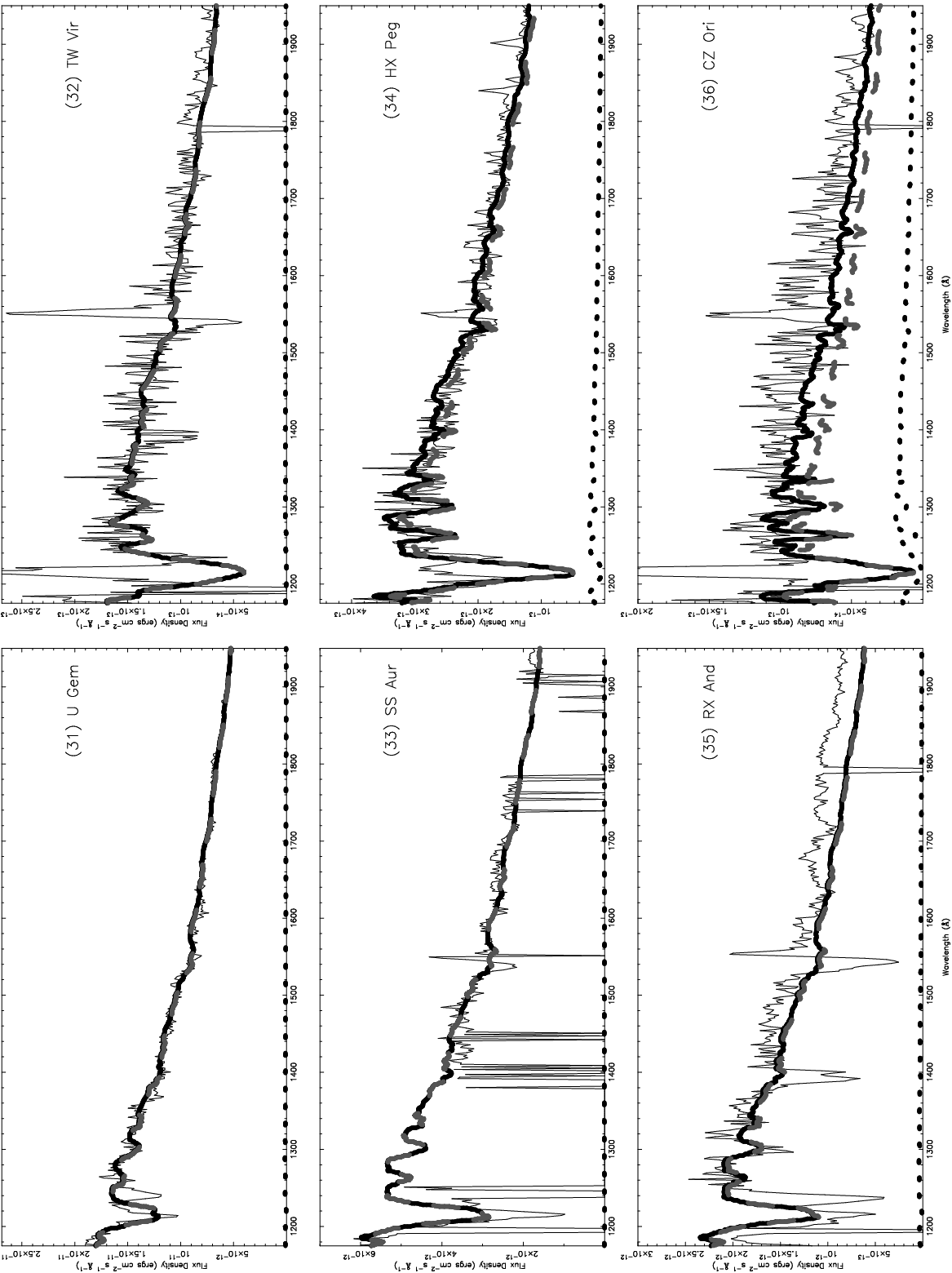}
\caption{Same as Figure 1 except for (31) U Gem, (32) TW Vir, (33) SS Aur, (34) HX Peg,
(35) RX And, (36) CZ Ori}
\end{figure}
\clearpage

\begin{figure}
\epsscale{0.92}
\plotone{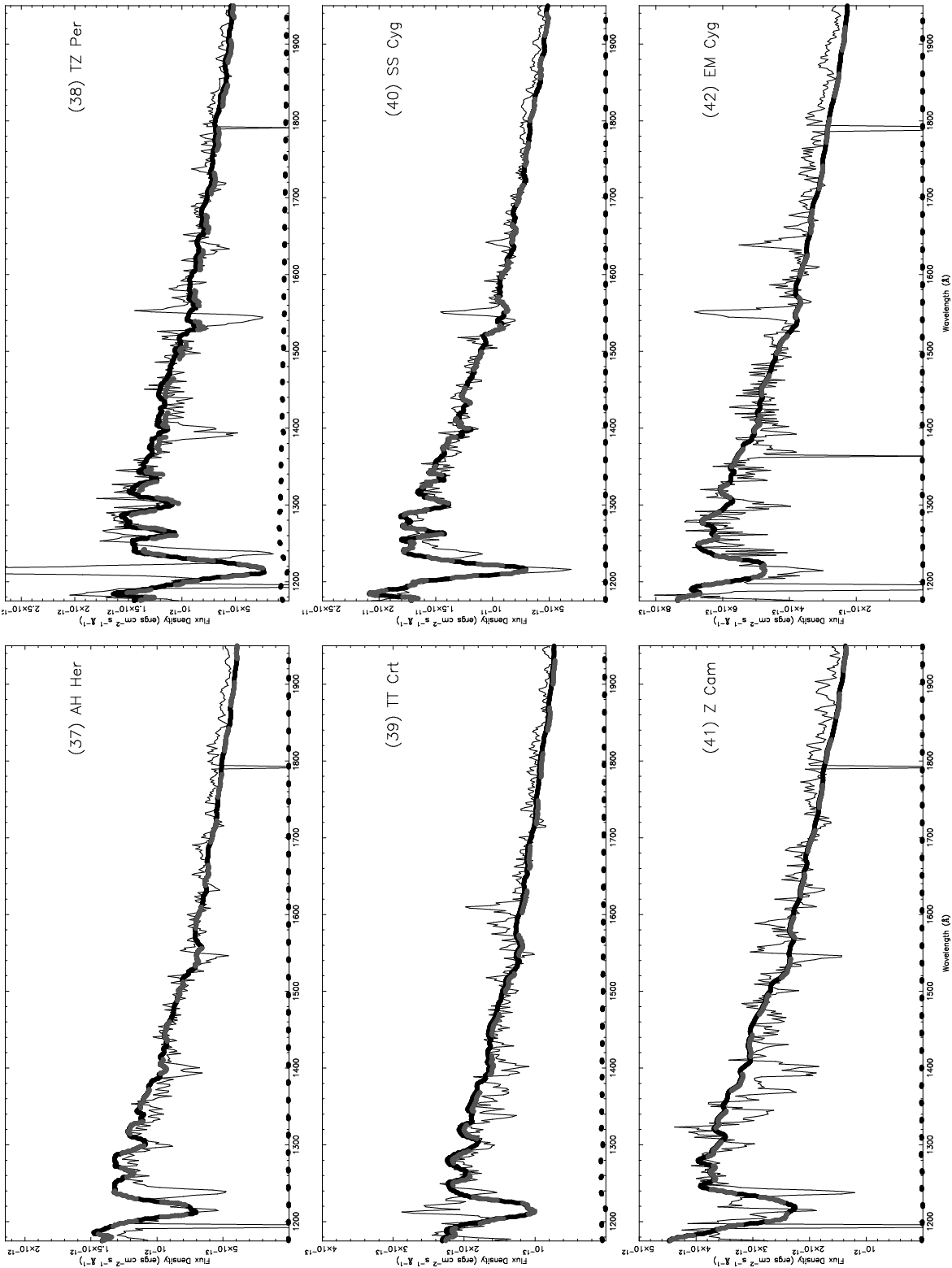}
\caption{Same as Figure 1 except for (37) AH Her, (38) TZ Per, (39) TT Crt, (40) SS Cyg,
(41) Z Cam, (42) EM Cyg}
\end{figure}
\clearpage

\begin{figure}
\epsscale{0.575}
\plotone{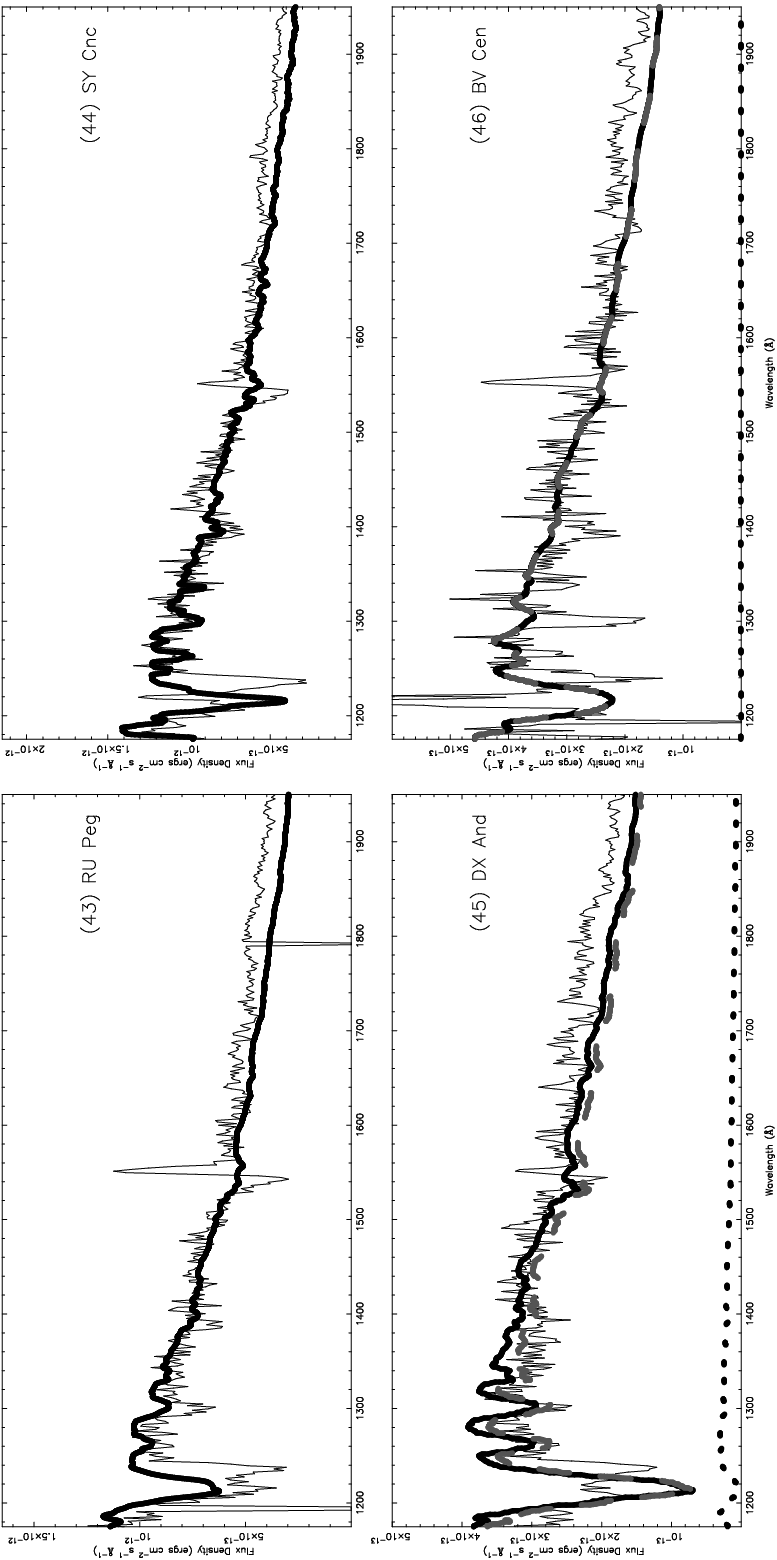}
\caption{Same as Figure 1 except for (43) RU Peg, (44) SY Cnc, (45) DX And, (46) BV Cen}
\end{figure}
\clearpage

\end{document}